\documentclass[aps,prb,preprint,groupedaddress]{revtex4}
\usepackage{graphicx} 
\usepackage{dcolumn}
\usepackage{bm}
\begin{document}

\title{Anomalous magnetic properties of multi-walled carbon nanotubes 
embedded with magnetic nanoparticles: Consistent with
ultrahigh temperature superconductivity} 
\author{Guo-meng Zhao$^{1,2,*}$, Jun Wang$^{2}$, Yang
Ren$^{3}$, and Pieder Beeli$^{1}$} 
\affiliation{
$^{1}$ Department of Physics and Astronomy, \\California State University at Los
Angeles,  Los Angeles, CA 90032,
USA~\\
\\$^{2}$Department of Physics, Faculty of
Science, Ningbo University, Ningbo, P. R. China~\\
\\$^{3}$X-Ray Science Division, Advance Photon
Source, Argonne National 
Laboratory, Argonne, IL 60439, USA}

 \begin{abstract}

We report high-temperature (300-1120 K) magnetization data of Fe 
and Fe$_{3}$O$_{4}$ nanoparticles embedded 
in multi-walled carbon nanotubes. The magnetic impurity concerntations
are precisely determined by both high-energy synchrotron x-ray
diffractometer and inductively coupled plasma mass spectrometer. We unambiguously show 
that the magnetic moments of
Fe and Fe$_{3}$O$_{4}$ nanoparticles are enhanced by a factor of about
3 compared with what they 
would be expected to have for free (unembedded) magnetic nanoparticles. 
The magnetization enhancement factor is nearly independent of the
applied magnetic field but depends significantly on the cooling rate. What is 
more intriguing is that the enhanced moments were completely lost when the 
sample was heated up to 1120 K and the lost moments at 1120 K were
completely recovered through several thermal cycles below 1020 K.
Furthermore, there is a rapid increase or decrease in the magnetization
below about 60 K. The anomalous magnetic properties 
cannot be explained by existing physics models except for the paramagnetic Meissner effect
due to the existence of ultrahigh temperature superconductivity in the multi-walled carbon nanotubes. 

\end{abstract}

\maketitle
\narrowtext
\section{Introduction}
Graphene is a sheet of carbon atoms distributed in a honeycomb lattice and is the building block 
for graphite and carbon nanotubes. Originating from conical valence 
and conduction bands that meet at a single point in momentum space, 
the massless charge carriers of graphene, known as Dirac fermions, exhibit 
relativistic behavior. Strong electron-electron
correlation of the Dirac fermions in graphene can lead to the formation of a short-range
resonating-valence-bond (RVB) liquid \cite{Meng} and/or to
a ferromagnetic instability \cite{Bas}. On the basis of the 
RVB theory of superconductivity originally proposed by Anderson
\cite{Anderson}, heavily doped graphene can 
exhibit ultrahigh temperature ($\sim$5000~K) $d$-wave superconductivity \cite{Black}.
 These theoretical works appear to agree with experimental
 observations of the intrinsic 
high-temperature ferromagnetism in graphite and carbon-based materials
\cite{Kop00,Maple,Mom,Esq03,Cer,Oh} and 
possible high-temperature superconductivity in carbon films \cite{Anto,Leb}, highly oriented 
pyrolithic graphite \cite{Kop00,Dus11}, carbon nanotubes
\cite{Wang,Zhaobook1,Zhaobook2}, and amorphous carbon \cite{Felner}. In addition, a giant 
magnetization enhancement has recently been identified in nickel nanoparticles embedded 
in multi-walled carbon nanotubes (MWCNTs) \cite{Zhao2010}. The giant 
magnetization enhancement was
tentatively explained in terms of ultrahigh temperature
superconductivity in MWCNTs \cite{Zhao2010}.  

Here we report high-temperature (300-1120 K) magnetization data of Fe 
and Fe$_{3}$O$_{4}$ nanoparticles embedded 
in multi-walled carbon nanotubes. Similar to the previous finding for nickel  
nanoparticles \cite{Zhao2010}, we unambiguously show that the magnetic moments of
Fe and Fe$_{3}$O$_{4}$ nanoparticles are also enhanced by a factor of about
3. The magnetization enhancement factor is nearly independent of the
applied magnetic field but depends significantly on the cooling rate. What is 
more intriguing is that the enhanced moments were completely lost when the 
sample was heated up to 1120 K and the lost moments at 1120 K were
completely recovered through several thermal cycles below 1020 K.
Furthermore, there is a rapid increase or decrease in the magnetization
below about 60 K. The anomalous magnetic properties 
cannot be explained by existing physics models except for the paramagnetic Meissner effect
due to the existence of ultrahigh temperature superconductivity in the multi-walled carbon nanotubes.

\section{Experiments} 
 
MWCNT mat samples embedded with Fe and Fe$_{3}$O$_{4}$ nanoparticles
 were obtained from SES Research of Houston (Catalog No. RS0657). The
 mat samples were synthesized by chemical vapor deposition under catalyzation of 
 Fe nanoparticles. During the purification process, some Fe nanoparticles 
 were oxidized into the Fe$_{3}$O$_{4}$ and $\alpha$-Fe$_{2}$O$_{3}$
 phases and were removed by HCl.  As confirmed by high-energy synchrotron 
x-ray diffraction (XRD) data (see Fig.~2 below), some fractions of Fe,
 Fe$_{3}$O$_{4}$, and $\alpha$-Fe$_{3}$O$_{4}$ nanoparticles still
 remain due to incomplete
 purification. The metal-based  impurity 
 concentrations of the mat sample were also determined by a Perkin-Elmer Elan-DRCe inductively 
 coupled plasma mass spectrometer (ICP-MS), which yielded the 
 metal-based magnetic impurity concentrations by weight: Fe = 
0.69$\%$, Co = 0.0036$\%$, Ni = 0.0021$\%$. 

\begin{figure}[htb]
    \centerline{\includegraphics[width=6.5cm]{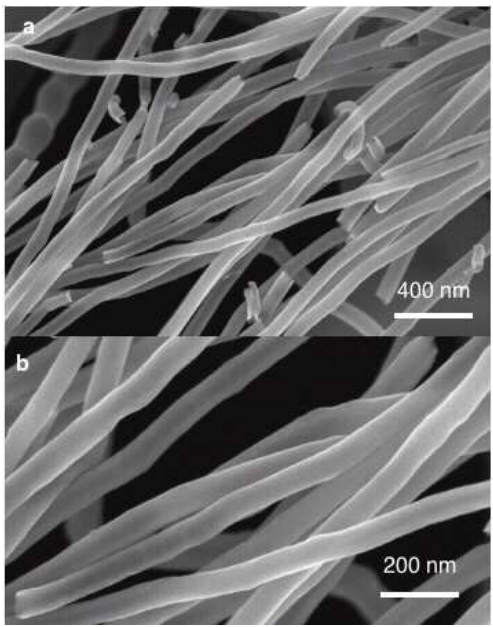}}
 \caption[~]{Scanning electron 
microscopy  (SEM) images of the MWCNT mat sample. The tubes are quite uniform and 
 their mean outer diameter is about 70 nm.}
\end{figure}

The morphology of the mat sample can be checked from scanning electron 
microscopy  (SEM) images shown in Fig.~1. The SEM images were taken by a field 
emission scanning 
electron microcopy (FE-SEM, Hitachi S-4800) using an accelerating voltage of 3 kV. 
One can see that the tubes are quite uniform and their mean outer diameter is about 70 nm.

\begin{figure}[htb]
    \centerline{\includegraphics[width=6.5cm]{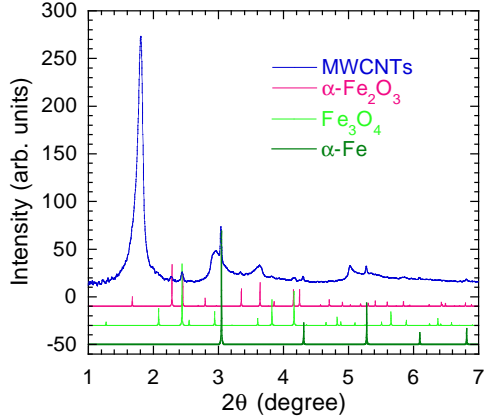}}
 \caption[~]{High-energy synchrotron x-ray diffraction spectrum of
 a virgin MWCNT mat sample along with the standard spectra of
$\alpha$-Fe, Fe$_{3}$O$_{4}$, and $\alpha$-Fe$_{3}$O$_{4}$ phases. Quantitative 
analyses of 
the XRD spectrum (see below) show that the sample 
contains  0.241$\pm$0.004$\%$ (by weight) of Fe, 0.216$\pm$0.015$\%$ of 
$\alpha$-Fe$_{2}$O$_{3}$, and 0.250$\pm$0.010$\%$ of Fe$_{3}$O$_{4}$
and that the mean 
diameters of Fe, $\alpha$-Fe$_{2}$O$_{3}$, and  Fe$_{3}$O$_{4}$ nanoparticles
are 46, 23, and 18~nm, respectively. }
 \label{fig2}
\end{figure}

Since the magnetic impurity phases in the MWCNT sample are so minor, it is impossible to
identify the minor phases from a normal low-energy x-ray diffraction (XRD) 
spectrum. But we can precisely determine magnetic impurity concentrations from 
high-energy synchrotron x-ray diffraction data \cite{Zhao2010}. Fig.~2 shows a synchrotron XRD 
spectrum for the MWCNT sample along with the standard spectra of
$\alpha$-Fe, Fe$_{3}$O$_{4}$, and $\alpha$-Fe$_{3}$O$_{4}$ phases. Using monochromated 
radiation with a wavelength of $\lambda$ = 0.1078~\AA, the XRD spectrum was 
taken on a high-energy synchrotron x-ray beam-line 11-ID-C at the Advanced Photon Source, Argonne National Laboratory. 
 In addition to the major peaks corresponding to 
the diffraction 
peaks of MWCNTs \cite{Rez}, there are many minor peaks, which match
well with all the peaks of  the $\alpha$-Fe, Fe$_{3}$O$_{4}$, 
and $\alpha$-Fe$_{2}$O$_{3}$ phases. This indicates that the visible impurity 
phases are $\alpha$-Fe, Fe$_{3}$O$_{4}$, and $\alpha$-Fe$_{2}$O$_{3}$,
in agreement with the impurity analysis using the ICP-MS above. 

In Fig.~3a, we display the expanded view 
of the MWCNT (002) peak for this sample.  The solid red line in Fig.~3a is the fitted curve by 
the sum of a Gaussian and a cut-off Lorentzian function, which takes into 
account both 
domain size broadening and 
strain broadening \cite{Rez}. The Lorentzian function is cut off to
zero when $|2\theta-2\theta_{\circ}|$ $\geq$ 3.65$\gamma$, where $2\theta_{\circ}$ is 
the peak
position and $\gamma$ is the full width at half maximum (FWHM). This
cut-off Lorentzian function plus a Gaussian function can excellently fit the MWCNT (002) peak
of Ref.~\cite{Rez}. 

\begin{figure}[htb]
    \centerline{\includegraphics[width=13cm]{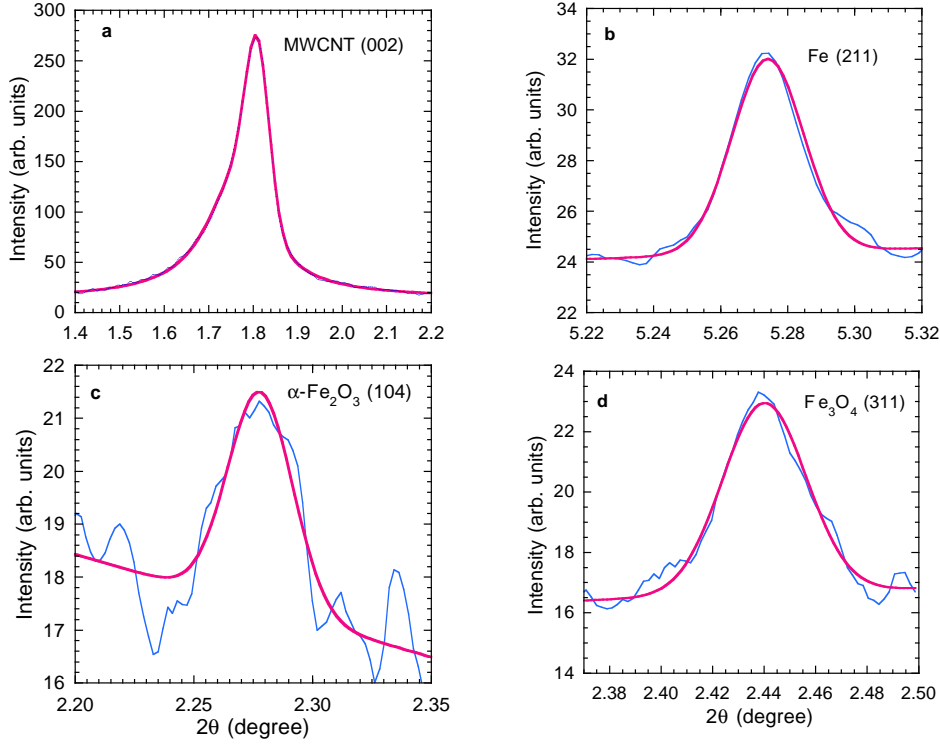}}
 \caption[~]{a) The expanded view of the MWCNT (002) peak. b) The expanded view of the Fe (211) peak. c) The 
 expanded view of the $\alpha$-Fe$_{2}$O$_{3}$ (104) peak. d) The expanded view of 
 the Fe$_{3}$O$_{4}$ (311) peak.}
\end{figure}

Figure~3b shows the expanded view 
of the Fe (211) peak of sample RS0657.
 The solid red line is the fitted curve by a Gaussian function. The Gaussian function 
is consistent with particle-size 
broadening \cite{Rez}. The 
integrated intensity of the Fe (211) peak is 
found to be
0.206$\pm$0.004$\%$ of the intensity of the 
MWCNT (002) peak. Using the standard intensities of graphite's (002) 
peak and Fe (211) peak and 
assuming that the intensity of MWCNT (002) peak is the same as that of 
graphite (002) peak, we find that the 
ferromagnetic Fe concentration is 0.241$\pm$0.004$\%$ (by weight).

Figures~3c and~3d display the expanded views 
of the $\alpha$-Fe$_{2}$O$_{3}$ (104) and Fe$_{3}$O$_{4}$ (311) peaks, respectively.  
The 
solid red line in Fig.~3c is the fitted curve by a Gaussian function. The 
integrated intensity of the $\alpha$-Fe$_{2}$O$_{3}$ (104) peak is 
found to be
0.348$\pm$0.024$\%$ of the intensity of the 
MWCNT (002) peak. From the intensity ratio, we find that the 
$\alpha$-Fe$_{2}$O$_{3}$ concentration is 0.216$\pm$0.015$\%$.

The spectrum of the Fe$_{3}$O$_{4}$ (311) peak in Fig.~3d
is obtained by subtracting the expected $\alpha$-Fe$_{2}$O$_{3}$ (110)
peak [whose integrated intensity is 76$\%$ of that for the 
$\alpha$-Fe$_{2}$O$_{3}$ (104) peak] from 
the raw spectrum. The solid red line in
Fig.~3d is the fitted curve by a Gaussian function. The 
integrated intensity of the Fe$_{3}$O$_{4}$ (311) peak is 
found to be
0.647$\pm$0.020$\%$ of the intensity of the 
MWCNT (002) peak. From the intensity ratio, we find that the 
Fe$_{3}$O$_{4}$ concentration is 0.250$\pm$0.010$\%$.

The quantitative analyses of 
the XRD spectrum show that the sample 
contains  0.241$\pm$0.004$\%$ of Fe, 0.216$\pm$0.015$\%$ of 
$\alpha$-Fe$_{2}$O$_{3}$, and 0.250$\pm$0.010$\%$ of 
Fe$_{3}$O$_{4}$. These impurity phases contribute to a metal-based Fe 
concentration of 0.58$\pm$0.02$\%$, which is about 0.11$\pm$0.04$\%$ lower than the total Fe
concentration (0.69$\pm$0.02$\%$) determined from the ICP-MS. Below we shall 
attribute this 0.11$\pm$0.04$\%$ discrepancy$-$which is not visible in the XRD spectrum
but can be clearly seen in magnetization data$-$to a minor phase of Fe$_{3}$C.

 We can determine mean diameters of magnetic nanoparticles from the 
 XRD peak widths.  For the Fe$_{3}$O$_{4}$ (311) peak, the full
width at half maximum $\gamma$ = 0.0382(1)$^{\circ}$. Using the Scherrer equation: 
$d = 0.89\lambda/(\gamma_{b} \cos \theta)$ and with $\gamma_{b}$ =
0.0312$^{\circ}$ (after correcting 
for the instrumental broadening $\gamma_{i}$ = 0.0221$^{\circ}$), we calculate 
$d$ = 18 nm. Similarly, the mean diameters of $\alpha$-Fe$_{2}$O$_{3}$ and  
Fe nanoparticles
are evaluated to be 23 and 46~nm, respectively.

\begin{figure}[htb]
    \centerline{\includegraphics[width=.4\textwidth]{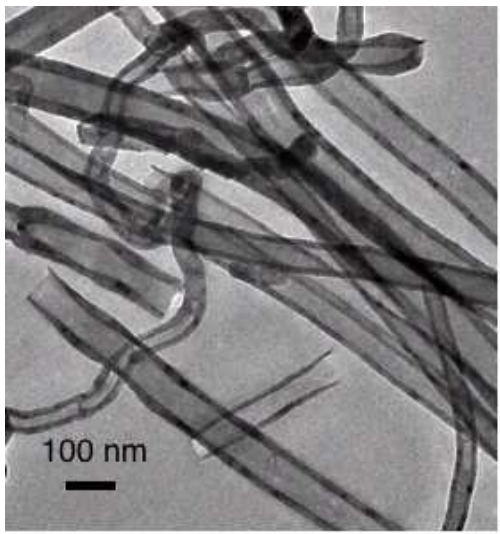}}
 \caption[~]{Transmission electron microscopic (TEM) image of the
 MWCNTs, which was recorded by FEI Tecnai F20 with an accelerating
voltage of 200 kV. }
\end{figure}

There are two ways to determine the mean inner diameter of the MWCNTs.
Scanning tunneling microscopic images shown in Fig.~1 indicate that the mean 
outer diameter of the tubes is about 70~nm.
Using the Scherrer equation, the mean wall thickness of the MWCNTs is 
determined to be about 9~nm from the FWHM value of the Gaussian function obtained 
by fitting the MWCNT (002) peak above. 
Therefore, the mean inner diameter of the MWCNTs is about 50~nm, close 
to the mean diameter of the Fe nanoparticles. This is consistent with 
the transmission electron microscopic (TEM) image shown in Fig.~4.
The TEM image also indicates that the mean wall thickness of the MWCNTs is about 
10~nm, in agreement with the XRD result.

Figure~5 shows zero-field-cooled (ZFC) and field-cooled (FC)
susceptibilities for a virgin sample of RS0657. Magnetic moment was measured 
using a Quantum Design vibrating 
sample magnetometer (VSM).  The absolute measurement uncertainties 
in temperature and moment are less than 20 K and 1$\times$10$^{-6}$
emu, respectively. The sample was first heated up to 1000 K and cooled
 down to 320~K in a ``zero'' ($<$0.06 Oe) field. A magnetic field of 
 0.5 Oe was applied at 320~K and the ZFC susceptibility was measured
 upon warming up to 1000~K. The FC susceptibility was taken when the
 temperature was lowered from 1000~K to 320~K. The FC and ZFC
 susceptibility data clearly show a magnetic transition at about 850~K, which 
 is associated with the
 ferrimagnetic transition of the Fe$_{3}$O$_{4}$ impurity phase. There
 is also a second magnetic transition at about 480~K, which 
 corresponds to  the ferromagnetic transition of the Fe$_{3}$C impurity phase.  
 
 \begin{figure}[htb]
    \centerline{\includegraphics[width=6.2cm]{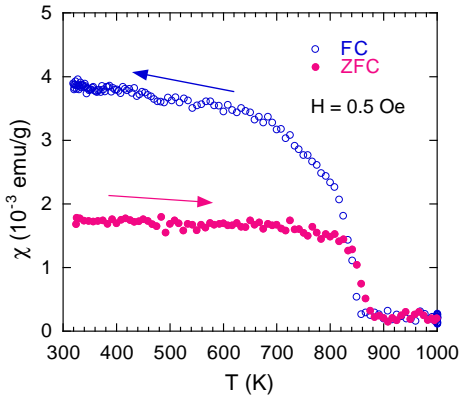}}
 \caption[~]{Zero-field-cooled (ZFC) and field-cooled (FC)
susceptibility data for a virgin MWCNT sample of RS0657. The FC and ZFC
 susceptibilities clearly show a magnetic transition at about 850 K, which 
 is associated with the
 ferrimagnetic transition of the Fe$_{3}$O$_{4}$ impurity phase. There
 is also a second magnetic transition at about 480~K, which 
 corresponds to the ferromagnetic transition of the Fe$_{3}$C impurity phase.}
 \label{fig5}
\end{figure}

Figure 6 shows magnetic hysteresis loops at 315 K and 910 K for 
sample RS0657, respectively.  The linear field dependence of the magnetization with a negative slope at 
$H >$ 10 kOe is due to the 
diamagnetic contribution. The linear extrapolation to $H$ = 0 yields 
$M_{s}$ = 1.66 emu/g at 315 K and $M_{s}$ = 0.62 emu/g at 910 K. The
temperature dependence of the saturation magnetization is displayed in
Fig.~7a. From the magnetic hysteresis loops, we also determine the coercive field
$H_{C}$, which is summarized in Fig.~7b. It is interesting that $H_{C}$ is
negligibly small
above 840 K, in agreement with a negligible remanent
magnetization above 840 K (see Fig.~8a). Another remarkable feature
is that the remament magnetization is almost reversible between 315
and 910 K (see Fig.~8b).

\begin{figure}[htb]
    \centerline{\includegraphics[width=7cm]{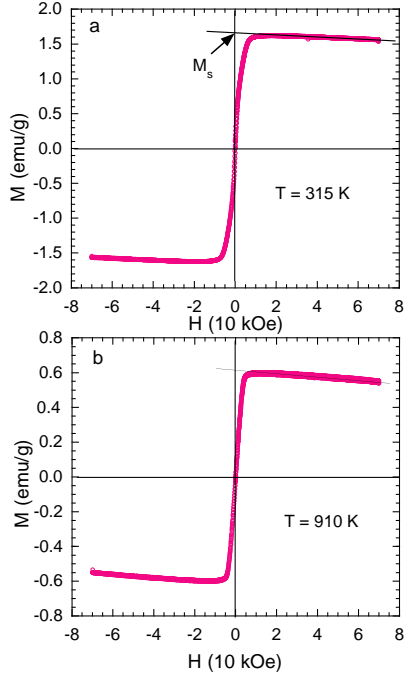}}
 \caption[~]{a) Magnetic hysteresis loop at 315 K for the MWCNT
sample. b) Magnetic hysteresis loop at 910 K for the MWCNT
sample. }
\end{figure}

From the hysteresis
loops, we can also determine the initial low-field susceptibility
$\chi_{i}$ from
the linear field dependence of the magnetization below 200 Oe. We find
that $\chi_{i}$ =  5.0$\times$10$^{-4}$ emu/g at 300 K and $\chi_{i}$
= 1.9 $\times$10$^{-4}$ emu/g at 910 K. It is known that the initial 
low-field susceptibility 
only depends on the demagnetization factor for strong ferromagnets
such as Fe and Fe$_{3}$O$_{4}$. For spherical particles with a demagnetization 
factor of 1/3, $\chi_{i}$ = 3/4$\pi$~emu/cm$^{3}$ =
0.239~emu/cm$^{3}$. At 910 K, only the Fe impurity phase contributes the 
initial susceptibility, which would be 7.35$\times$10$^{-5}$ emu/g if 
the 0.24$\%$ Fe nanoparticles were decoupled from  the MWCNTs.
Therefore, the initial susceptibility of the embedded Fe nanoparticles
is enhanced by a factor of 2.6 compared with what they would be
expected to have if they would be isolated from the MWCNTs. For
Fe$_{3}$O$_{4}$ nanoparticles with the concentration of 0.25$\%$, the expected initial susceptibility would be
1.2$\times$10$^{-4}$ emu/g, which is also a factor of 2.6 smaller than the measured
difference (3.1$\times$10$^{-4}$ emu/g) between the susceptibilities at 300 K and 910~K.

\begin{figure}[htb]
    \centerline{\includegraphics[width=7cm]{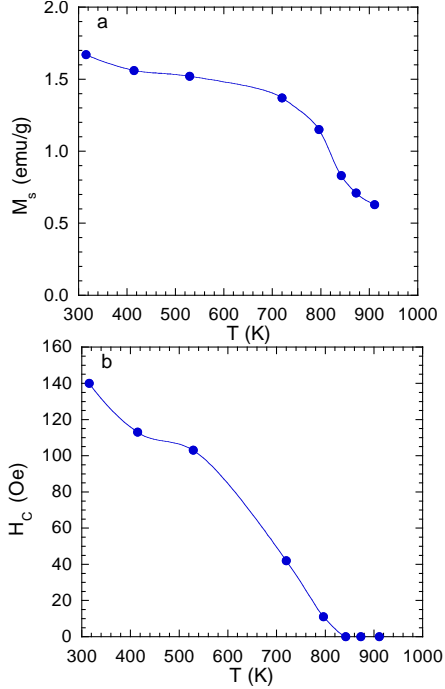}}
 \caption[~]{a) Temperature
dependence of the saturation magnetization $M_{s}$. b) Temperature
dependence of the coercive field $H_{C}$.   }
\end{figure}

Figure 9a shows temperature dependencies of the high-field (10 kOe)
magnetizations for
another virgin MWCNT sample of RS0657.  Since the magnetization in 10 kOe is close to the saturation
magnetization (see Fig.~6), the temperature dependence of the
saturation magnetization is approximated by the temperature
dependence of the magnetization in 10 kOe. The first warm-up magnetization 
curve clearly demonstrates three magnetic transitions at about 500 K, 860 K,
and 1060 K, which should be associated with the 
phases of the magnetic Fe$_{3}$C, Fe$_{3}$O$_{4}$, and $\alpha$-Fe ($T_{C}$ = 1047 K), respectively.
The slightly higher magnetic transition temperatures than the expected
ones are due to a thermal lag.

\begin{figure}[htb]
    \centerline{\includegraphics[width=7cm]{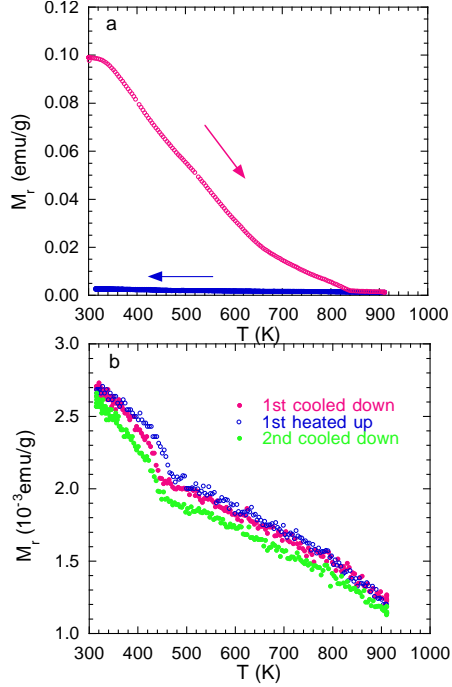}}
 \caption[~]{a) Temperature dependence of the initial remanent
magnetization. The remanent magnetization was measured in a field of less than 0.03 Oe after the
sample was magnetized in a magnetic field of 70 kOe at 300 K. b)
Thermal hysteresis of the remanent magnetization between 315 and 910 K.  }
\end{figure}

One can clearly see that the thermal hysteresis of the high-field magnetization is anomalous. When the sample was 
cooled from 1120 K, the cool-down magnetization was reduced dramatically compared with the first 
warm-up magnetization. After two more thermal cycles below 1020 K, the 
lost magnetization at 1120~K was recovered almost completely.

In Fig.~9b, we decompose the high-field magnetization ($\simeq$ $M_{s}$) into 
contributions
from the $\alpha$-Fe and Fe$_{3}$O$_{4}$ impurity phases on the assumption that
$M_{s}(t)/M_{s}(0)$  
for both Fe and Fe$_{3}$O$_{4}$ impurity 
phases is the same as that for free (unembedded) $\alpha$-Fe nanoparticles (where
$t$ = $T/T_{C}$). The $M_{s}(t)/M_{s}(0)$ curve for unembedded $\alpha$-Fe nanoparticles
with the mean diameter of over 100 nm was also measured in a field of 
10 kOe. The solid 
blue line is the contribution from the Fe impurity phase with
$M_{s}(300 K)$ = 0.96 emu/g  and the solid green line is the contribution
from the Fe$_{3}$O$_{4}$ impurity phase with
$M_{s}(300 K)$ = 0.46 emu/g. The remaining $M_{s}(300 K)$ = 0.18 emu/g
should contribute from the Fe$_{3}$C impurity phase.

For unembedded Fe nanoparticles with a mean
diameter of 46 nm, the saturation magnetization can be extrapolated to be 
160~emu per gram of 
Fe from the measured diameter
dependence of $M_{s}(300 K)$ (Ref.~\cite{Gong}). Using the
Fe concentration of 0.24$\%$, $M_{s}(300 K)$ is calculated to be 0.38 emu/g
if the Fe nanoparticles would be isolated from the MWCNTs. This clearly
indicates that the saturation magnetization (0.96 emu/g) of 
the 46-nm Fe
nanoparticles embedded in MWCNTs is enhanced by a factor of about 2.5 
compared with that (0.38 emu/g) of the unembedded Fe nanoparticles. This magnetization 
enhancement 
factor is very close to that (2.6) found above from the initial
low-field magnetization. This indicates that the magnetization enhancement
factor is nearly independent of the magnetic field.

\begin{figure}[htb]
    \centerline{\includegraphics[width=6.7cm]{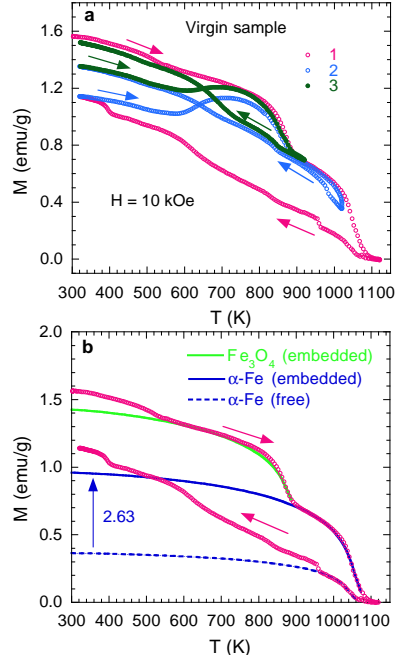}}
 \caption[~]{  a)  Temperature dependencies of the high-field
 magnetizations for
 a virgin MWCNT sample of RS0657, which were measured in a magnetic field of 10
 kOe. There are three thermal cycles, labeled by 1, 2 and 3.
 b) Magnetic
decomposition into contributions from the $\alpha$-Fe and Fe$_{3}$O$_{4}$ impurity 
phases. The solid 
blue and green lines are the contribution from the Fe and Fe$_{3}$O$_{4}$
impurity phases, respectively. The dashed blue line is the expected
contribution from 
the unembedded 46-nm Fe nanoparticles with the 
concentration of 0.23$\%$ (very close to 0.241$\pm$0.004$\%$
determined from the XRD data).  }
 \label{fig9}
\end{figure}

For unembedded Fe$_{3}$O$_{4}$ nanoparticles 
with a mean diameter of 18 nm, the $M_{s}(300 K)$ value can be
inferred to be about   
62~emu per gram of Fe$_{3}$O$_{4}$ from the measured diameter
dependence of $M_{s}(300 K)$ (Ref.~\cite{Goya}). With the Fe$_{3}$O$_{4}$
concentration of 0.25$\%$,  we calculate $M_{s}(300 K)$ to be 0.155 
emu/g if the Fe$_{3}$O$_{4}$ nanoparticles would be isolated from the MWCNTs.
It is clear that the $M_{s}(300 K)$ value (0.46 emu/g) of the 
18-nm Fe$_{3}$O$_{4}$ nanoparticles embedded in MWCNTs is enhanced by a factor of about 3.0 
compared with that (0.155 emu/g) of the unembedded Fe$_{3}$O$_{4}$ nanoparticles.  This
magnetization enhancement 
factor is similar to that for 11-nm nickel nanoparticles \cite{Zhao2010}.

The dashed blue line in Fig.~9b is the expected magnetization
curve for unembedded Fe nanoparticles with 
the mean diameter of 46 nm and the 
concentration of 0.23$\%$ (very close to 0.241$\pm$0.004$\%$
determined from the XRD data). This curve matches very 
well with the first 
cool-down magnetization data between 960 and 1060~K. This implies that the
enhanced magnetization was completely lost at 1120~K and not
recovered upon cooling down to 960~K. This result vividly
demonstrates that the Fe impurity concentration determined from the
XRD data is precise. By comparing the solid blue line to the dashed blue
line, we find the magnetization enhancement 
factor to be 2.63. This parameter-free
determination of the magnetization enhancement factor justifies our
XRD analyses and provides indisputable evidence for the giant
magnetization enhancement of magnetic nanoparticles embedded in
MWCNTs.

It is interesting that in the first cool-down curve there 
are two sharp increases in the magnetization at about 960 K and 400 K. There 
are also two gradual increases at about 860 K and 660 K. The small
magnetization increase (about 0.1 emu/g) at 860 K would 
be expected for the ferrimagnetic transition of the Fe$_{3}$O$_{4}$
impurity phase if they would be isolated from the MWCNTs.  This suggests that the enhanced magnetization of 
the Fe$_{3}$O$_{4}$
impurity phase seen in the first warm-up 
magnetization data was also lost upon heating up to 1120 K. In fact,
the enhanced magnetization of the Fe$_{3}$O$_{4}$
impurity phase was lost almost completely at 920~K, as seen from curve
3 in Fig.~9a. The lost
magnetization of the Fe phase was partially 
gained back at 960 K. The lost magnetizations of both Fe$_{3}$O$_{4}$ 
and Fe phases were completely recovered through 
two more thermal cycles below 1020 K (see Fig.~9a). It is striking that$-$independent of 
whether the sample is heated or cooled$-$the significant magnetization gain 
always occurs in the temperature region between 600 and 700 K.

\begin{figure}[htb]
    \centerline{\includegraphics[width=7cm]{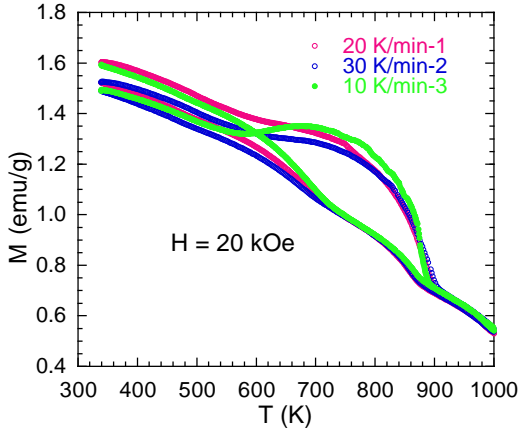}}
 \caption[~]{Temperature dependencies of the magnetizations measured with different
heating/cooling rates.}
 \label{fig10}
\end{figure}

In Fig.~10, we plot the temperature dependencies of the magnetizations,
which were measured in a magnetic field of 20 kOe and with different
heating/cooling rates. Fig.~10 indicates
that the cool-down magnetizations below 720 K depend significantly on 
the cooling rates; the slower the cooling rate is, the larger the
magnetization is. The cooling-rate dependence of the cool-down
magnetization is another unusual magnetic property of our samples. 

Figure~11a shows low-temperature susceptibility curves measured in a
magnetic field of 1 kOe. From the first cool-down curve, one 
can see that the susceptibility drops sharply at $T_{on}$ = 40 K, reaches a
minimum value at $T_{d}$ = 28 K, and increases rapidly as the temperature is
further reduced. Upon warming, the susceptibility does not follow the 
cool-down curve; both $T_{on}$ and $T_{d}$ shift to higher
temperatures and the susceptibility value at $T_{d}$ becomes much
smaller. One more thermal cycle leads to a small increase in $T_{on}$ 
and a large drop in the susceptibility below 40 K. Furthermore, the
susceptibilities below $T_{on}$ show a large fluctuation, suggesting a 
metastable magnetic state.

\begin{figure}[htb]
    \centerline{\includegraphics[width=7cm]{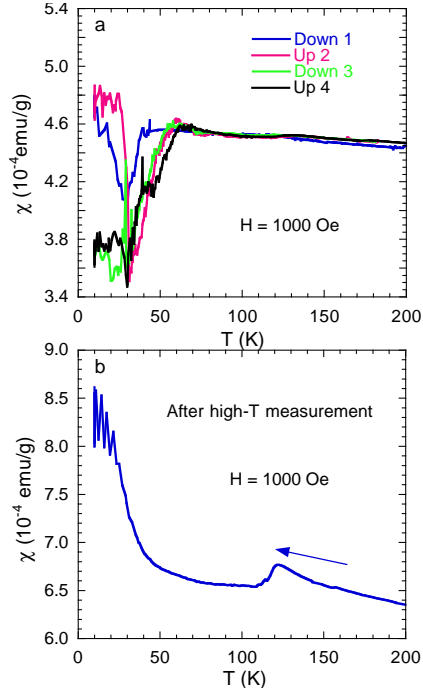}}
 \caption[~]{a) Low-temperature susceptibility curves measured in a
magnetic magnetic field of 1 kOe. b) Low-temperature susceptibility curve
after high-temperature susceptibility data were taken.}
 \label{fig11}
\end{figure}

It is remarkable that low-temperature magnetic behavior changed
dramatically after high-temperature susceptibility measurement were
done (see Fig.~11b). The susceptibility do not drops but increases sharply 
at $T_{on}$. A sharp drop at about 122~K is associated with
the Verwey transition of the Fe$_{3}$O$_{4}$ impurity phase.

\section{Discussion}

Now we discuss possible explanations to the oberved anomalous magnetic properties of multi-walled carbon nanotubes 
embedded with magnetic nanoparticles. Firstly, we would like to check 
if the magnetic impurity phases identified from the high-energy XRD
spectrum and ICP-MS can explain the data. It is  known that high-field 
magnetizations for
ferromagnetic/ferrimagnetic materials have negligible thermal hysteresis
due to the fact that the magnetic field is large enough to saturate
the magnetization. Indeed, we have 
demonstrated that the high-field (10 kOe) magnetization of pure Fe$_{3}$O$_{4}$ 
nanoparticles has a 
negligible thermal hysteresis \cite{Wang2011}. Therefore, 
the observed anomalous thermal hysteresis of the high-field magnetization
cannot be explained by ferrimagnetic thermal hysteresis of Fe$_{3}$O$_{4}$ 
nanoparticles or ferromagnetic thermal hysteresis of Fe nanoparticles.

If we assume that the spin order of Fe$_{3}$O$_{4}$ nanoparticles  is 
not independent of the spin order of Fe nanoparticles,  the frustration effect 
arising from the competition between the interactions of the two types of
nanoparticles may lead to the aging effect, memory, and rejuvenation effect, which 
may account for the anomalous thermal hysteresis in the total magnetization.
If this scenario could explain the anomalous thermal hysteresis 
below the Curie temperature (850 K) of the Fe$_{3}$O$_{4}$ impurity phase, it 
could not consistently explain the anomalous thermal hysteresis above 850 K 
where Fe$_{3}$O$_{4}$ is paramagnetic and Fe is the only ferromagnetic phase. 
Furthermore, this scenario cannot consistently explain the giant moment 
enhancement. This is because the saturation magnetization of an 
Fe-core/Fe$_{3}$O$_{4}$-shell particle is a simple superposition of the 
$M_{s}$ contributions from Fe and Fe$_{3}$O$_{4}$ (see
Ref.~\cite{Gango}).

Another possibility is that the magnetism of Fe$_{3}$O$_{4}$ 
nanoparticles may be related to the superparamagnetism. In this case, 
all the spins in 
each nanoparticle are ordered at high temperatures. With decreasing temperature the spin 
direction of nanoparticles may be frozen, forming either a superparamagnet or 
a superspin glass. The anomalous memory effect due to the spin frustration may 
be associated with the anomalous thermal hysteresis. This possibility 
is unlikely because of the following reasons. Firstly, our Fe$_{3}$O$_{4}$ 
nanoparticles are not superparamagnetic because the coercive field and
remanent magnetization are nonzero before the Curie temperature is
reached, as seen from
Fig.~7b and Fig.~8a. Secondly, it is known that the saturation magnetization of the
superparamagnetic nanoparticles is always reduced compared with the
bulk value, in contrast to the giant enhancement. Thirdly, our magnetic data for 
pure 12-nm Fe$_{3}$O$_{4}$ nanoparticles \cite{Wang2011} do not show
sizable thermal hysteresis.

Since the identified magnetic impurity phases cannot explain the data,
we then shall consider interplay between the magnetic nanoparticles and
the MWCNTs. We have shown that \cite{Zhao2010} the giant magnetic moment 
enhancement of 
nickel nanoparticles embedded in MWCNTs is unlikely to arise from the magnetic 
proximity effect~\cite{Ces}. For the case 
of 11-nm nickel nanoparticles embedded in MWCNTs \cite{Zhao2010}, the 
induced moment $m$ per contact
carbon within this magnetic proximity model was calculated to be 5.1~$\mu_{B}$ (where $\mu_{B}$
is Bohr magneton). 
Similarly we calculate $m$ = 61 $\mu_{B}$ for 46-nm Fe nanoparticles embedded in MWCNTs
and $m$ = 8.0~$\mu_{B}$ for 18-nm Fe$_{3}$O$_{4}$ nanoparticles embedded in MWCNTs. Within the density 
function theory
\cite{Ces}, the induced moment per contact carbon 
is calculated to be about 0.1~$\mu_{B}$, which is too small to explain
the observed giant moment enhancements. Another difficulty with this scenario is that
it cannot explain the anomalous thermal hysteresis of the 
high-field magnetization and no observable changes in the Curie
temperatures of the nanoparticles.

It is possible that a strong diamagnetic tube could
enhance the extrinsic magnetic moment of a (single-domain) magnet
embedded inside it. If the tube were a perfect diamagnet, the ``poles'' 
of the magnet would be extended further apart (to the length of the tube) without
changing their strength, thus giving an extrinsic enhancement to the
magnetic moment. This is because the perfect diamagnetism of the tube 
prevents the magnetic field lines of the magnet from leaking out
through the wall of the tube. The plausibility of this interpretation
depends on whether MWCNTs would exhibit strong diamagnetism when the magnetic field is
applied in the tube-axis
direction. Since the orbital diamagnetism for the field parallel to the 
tube-axis direction is negligibly small, the strong diamagnetism could arise 
from superconductivity or ideal conductivity due to ballistic
transport. 

If the strong diamagnetism along the tube axes exists in the MWCNTs due to ballistic 
transport, the magnetizations of
the magnetic nanoparticles embedded inside innermost shells of the
MWCNTs can be enhanced when the magnetic field is applied at
a temperature below a criticial temperature $T_{b}$ where the ballistic
transport disappears.  The enhanced magnetizations should get lost above
$T_{b}$ due to the disappearance of the strong diamagnetism. Curves 3 and 1 in Fig.~9a
imply that 
the enhanced magnetization of
the Fe$_{3}$O$_{4}$ nanoparticles gets lost at 920 K while the
enhanced magnetization of the Fe nanoparticles remains at this
temperature and gets lost at 1120 K. This would imply two distinctive
$T_{b}$'s in the MWCNTs, which happen to coincide with the Curie temperatures of
the Fe$_{3}$O$_{4}$ and Fe phases. This is unlikely. More serious
problem with this scenario is that,  unlike a superconductor, there is no diamagnetism and 
thus no magnetization
enhancement when the sample is cooled down from a
temperature above $T_{b}$. This is in sharp contrast to what we have
observed. 

If the strong diamagnetism along the tube axes exists in the MWCNTs due to
superconductivity, the magnetization enhancement is always possible
below the superconducting transition temperature $T_{c}$ independent
of whether the field is applied at a temperature above or below
$T_{c}$. In order for this interpretion to be valid, one must assume
that there would be two distinctive
$T_{c}$'s in the MWCNTs, which are close to the Curie temperatures of
the Fe$_{3}$O$_{4}$ and Fe phases. This is also unlikely.
Therefore, the diamagnetic Meissner effect in superconducting MWCNTs should be very weak. Indeed, 
the small diamagnetic Meissner
effect along the tube axes has been identified for pure MWCNTs and the
magnitude of the diamagnetic Meissner effect is found to be in
quantitative agreement with the inferred magnetic penetration depth from
the measured carrier concentration \cite{Zhaobook2}. The very weak diamagnetic Meissner
effect in the MWCNTs is simply because the magnetic penetration depth
is significantly larger than the wall thicknesses of the tubes.

Finally, the giant magnetization enhancement, the anomalous thermal 
hysteresis of the high-field magnetization, and the unusual cooling-rate 
dependence of the high-field magnetization can be naturally explained 
in terms of the high-field paramagnetic Meissner
effect (HFPME) due to the existence of ultrahigh temperature  superconductivity in 
the MWCNTs. The HFPME has been 
observed in large crystals \cite{Kus} 
and melt-textured samples \cite{Dias} of YBa$_{2}$Cu$_{3}$O$_{7-y}$,
which was attributed to field-induced 
flux pinning (a ``fishtail'' effect) \cite{Kus}. The field-induced
flux pinning can be enhanced by
inclusion of magnetic impurity phases, as in the case of the melt-textured 
samples \cite{Dias} of YBa$_{2}$Cu$_{3}$O$_{7-y}$, which contain a paramagnetic 
impurity phase of Y$_{2}$BaCuO$_{5}$. The HFPME causes
an apparent increase of the paramagnetic magnetization of the impurity
phase (due to flux trapping on the paramagnetic impurity sites) and the magnetization 
enhancement factor is nearly independent
of the magnetic field (see Fig.~4 of Ref.~\cite{Dias}).  Remarkably,
the field-cooled magnetization at a fixed temperature was found to increase with
time \cite{Dias}, implying that the field-cooled magnetization should increase with
cooling rate. This is similar to what we have observed in our MWCNTs
embedded with ferromagnetic nanoparticles (see Fig.~10). 
Furthermore, curves 1 and 3 in Fig.~9a also demonstrate that the large
magnetization enhancements only occur at temperatures well below the Curie 
temperatures of the ferromagnetic impurity phases and the enhanced
magnetizations get lost above the Curie temperatures (850 K and 1050 K) of the Fe$_{3}$O$_{4}$
and Fe phases, respectively. This is 
consistent with the fact that
ferromagnetically ordered  impurities are very effective flux
pinning centers \cite{Riz} and that strong flux pinning causes the HFPME
\cite{Kus}. Therefore, all these unusual magnetic
properties can be well explained by the HFPME.

If individual MWCNTs are superconductors, a MWCNT mat should be
a granular superconductor. Below a Josephson-coupling temperature, the
magnetic response could be paramagnetic or diamagnetic depending on
the Josephson-coupling strength and magnetic field \cite{Chan}. The
data shown in Fig.~11 are consistent with the expected magnetic response of a granular superconductor.

If the HFPME interpretation is relevant, one must assume that
our MWCNTs should exhibit ultrahigh temperature superconductivity with $T_{c}$ $>$ 1050 K. This 
is consistent 
with a single-particle 
tunneling gap of about 200 meV, which was identified \cite{Zhaobook2} 
in a 30-nm MWCNT. Electrical transport data for a
MWCNT mat sample were also found to agree with granular
superconductivity below the mean-field $T_{c}$ of about 700 K.  
Further, the inelastic scattering rate at room temperature was 
found to be very large (about
30 meV) in graphite \cite{Sug}. With the Fermi-velocity of about
8$\times$10$^{5}$ m/s, one calculates the inelastic mean-free path 
 to be 18 nm in graphite, which is far shorter than that
(about 1 $\mu$m) estimated from electrical transport measurement
\cite{Dus}. For 
individual MWCNTs, the room-temperature mean-free path was determined 
to be larger 
than 65 $\mu$m (Ref.~\cite{Heer}). These transport data cannot be explained 
by ballistic transport but are consistent with phase-incoherent room-temperature
superconductivity.

\section{Concluding remark}
Because there have
been no traditional 
signatures of bulk superconductivity (e.g., large
diamagnetic Meissner effect and the zero-resistance state) in these
materials, our interpretation in terms of the HFPME might be
questionable. However, since we are not aware of any better
explanations based on other physics models, the
interpretion based on the HFPME 
due to ultrahigh temperature superconductivity in MWCNTs should be
the most relevant.

{\bf Acknowledgment:} We thank M. Du and F. M. Zhou for the 
elemental analyses using ICP-MS. Use of the Advanced Photon Source was supported 
by the U.S. Department of Energy, Office of 
Science, Office of Basic Energy Sciences, under Contract No. DE-AC02-06CH11357.
This work was partly supported by the National Natural Science Foundation of China (10874095)
and Y. G. Bao's Foundation.~\\
~\\

$^{*}$ Corresponding should be addressed to gzhao2@calstatela.edu

\end{document}